\documentclass{jps-cp}
\usepackage{txfonts} 

\usepackage{color}

\newcommand{\black}{\color{black}}

\title{Nucleon isovector tensor charge from lattice QCD with physical light quarks}

\author{
Ryutaro \textsc{Tsuji}$^{1.2}$, 
Yasumichi \textsc{Aoki}$^{2}$, 
Ken-Ichi \textsc{Ishikawa}$^{3}$, 
Yoshinobu \textsc{Kuramashi}$^{4}$, 
Shoichi \textsc{Sasaki}$^{1}$, 
Eigo \textsc{Shintani}$^{4}$ and
Takeshi \textsc{Yamazaki}$^{4.5}$\\
{\normalsize{\bf \sffamily \hspace{50mm}(PACS Collaboration)}}
}

\inst{
$^{1}$Department of Physics, Tohoku University, Sendai 980-8578, Japan \\
$^{2}$RIKEN Center for Computational Science, Kobe 650-0047, Japan\\
$^{3}$Core of Research for the Energetic Universe, Graduate School of Advanced Science and Engineering, Higashi-Hiroshima 739-8526, Japan\\
$^{4}$Center for Computational Science, University of Tsukuba, Tsukuba 305-8577, Japan\\
$^{5}$Faculty of pure and Applied Science, University of Tsukuba, Tsukuba 305-8571, Japan\\
}

\email{tsuji@nucl.phys.tohoku.ac.jp}

\recdate{January, 17, 2022}

\abst{
We present preliminary results for
the axial, scalar and tensor charges of the nucleon
measured in 2+1 flavor QCD with the physical light quarks ($m_\pi=135$ MeV).
Our simulations are carried out with gauge configurations generated by the PACS Collaboration with the stout-smeared $O(a)$ improved Wilson fermions and Iwasaki gauge action at a single lattice spacing of $0.085\ (\mathrm{fm})$.
There are two lattice ensembles of the PACS gauge configurations, which have physical lattice sizes
over $(10\ \mathrm{fm})^4$ and $(5\ \mathrm{fm})^4$, respectively.
We compute the nucleon three-point correlation functions in the axial, scalar, and tensor channels. 
For the renormalization, we use the Rome-Southampton method as the intermediate scheme 
in order to evaluate the renormalization constants for the scalar and tensor currents 
in fully nonperturbative manner.
We then evaluate the renormalized values of the scalar and tensor charges ($g_S$ and $g_T$)
in the $\overline{\rm MS}$ scheme at the renormalization scale of 2 GeV
with a help of the continuum perturbation theory for the matching 
between two schemes. We compare our preliminary results of $g_S$ and $g_T$
with those of other collaboration results.
}

\kword{Nucleon structure, Lattice QCD,  \ldots}

\begin{document}
\maketitle

\section{Introduction}
In the standard model of modern particle physics, the nucleon is known
to be a composite particle made of quarks and gluons, and their interactions are 
described by QCD. 
Due to the nonperturbative nature of QCD at low energy scales, 
the nucleon structure that is governed by strong many body problem 
of the elementary constituents is one of the great challenges of lattice QCD.
Future and current precision $\beta$-decay measurements with cold
and ultracold neutrons provides us an opportunity to study the
sensitivity of the nucleon isovector matrix elements to new physics
beyond the standard model (BSM). The neutron life-time puzzle is
one of such examples~\cite{Czaenecki2018}. The discrepancy between the results of
beam experiments and storage experiments remains unsolved. 
It is still an open question that deserves further investigation in 
terms of the nucleon axial charge ($g_A$). 
Although the nucleon axial charge dominates the weak decay of the neutron, there is
no reason to forbid the other channel contributions to the neutron decay
if the BSM contributions are present. In this sense, the scalar and tensor charges ($g_S$ and $g_T$),
which are less known experimentally, play important roles to constrain the limit
of non-standard interactions~\cite{Bhattacharya:2011qm,Courtoy2015,Vincenzo2013}.



\section{Method}
\label{sec:method}
The axial, tensor and scalar charges can be evaluated from the nucleon matrix element of a given bilinear operator, $O_{\Gamma}=\bar{\psi}\Gamma\psi$ with $\Gamma = \gamma_5\gamma_{i},\ 1$ and $\gamma_i\gamma_j(i\neq j)$, respectively.

In general, the nucleon matrix elements are evaluated from a ratio of the nucleon three-point function with a given operator $O_{\Gamma}$ inserted at $t=t_{\rm op}$ being subject to a range of $t_{\rm snk} > t > t_{\rm src}$, 
to the nucleon two-point function with a source-sink separation ($t_{\rm sep}= t_{\rm snk}-t_{\rm src}$) as 
\begin{align}
    \label{eq:method_ratio}
    R(t_{\rm sep},t_{\rm op})
    \equiv
    \frac{C^{\rm 3pt}(t_{\rm op},t_{\rm sep})}{C^{\rm 2pt}(t_{\rm sep})}
    \underset{t_{\rm sep}\gg t_{\rm op}\gg 0}{\longrightarrow}
    \langle 1 | O_{\Gamma} |1 \rangle
    + {\mathcal{O}}(\mathrm{e}^{-t_{\rm sep}\Delta})
    + {\mathcal{O}}(\mathrm{e}^{-(t_{\rm sep}-(t_{\rm op}-t_{\rm src})\Delta}),
\end{align}
where $|i\rangle$ represents the $i$-th energy eigenstate and $i=1$ stands for the ground state of the nucleon.  
If the condition $t_{\rm sep}\gg t_{\rm op}-t_{\rm src}\gg 0$
is satisfied, the desired matrix element $\langle 1 | O_{\Gamma} |1 \rangle$ can be read off from an asymptotic plateau, 
which is independent of a choice of $t_{\rm op}$.
Narrower source-sink separation causes
systematic uncertainties stemming from the excited states contamination represented by two terms of ${\mathcal{O}}(\mathrm{e}^{-t_{\rm sep}\Delta})$ and ${\mathcal{O}}(\mathrm{e}^{-(t_{\rm sep}-(t_{\rm op}-t_{\rm src})\Delta})$, where $\Delta\equiv E_{2}-E_{1}$ denotes 
a difference between the two energies
of the ground state ($E_1$) and the lowest excited state ($E_2$).

    In this study, the nucleon interpolating operator is constructed by
    the exponentially smeared quark operators, so as to maximize an
    overlap with the nucleon ground state as

\begin{align}
    \label{eq:interpolating_op}
    N(t,\vec{p})&
    =
    \sum_{\vec{x}\vec{x_1}\vec{x_2}\vec{x_3}}
    \mathrm{e}^{-i\vec{p}\cdot\vec{x}}\varepsilon_{abc}
    \left[
        u^{T}_{a}(t,\vec{x_1})C\gamma_5d_b(t,\vec{x_2})
    \right]
    u_c(t,\vec{x_3})
    \times 
    \Pi_{i=1}^{3}
    A\mathrm{exp}(-B|\vec{x_i}-\vec{x}|)
\end{align}
where there are two smearing parameters $(A, B)$. Since the condition, $t_{\mathrm{sep}}\gg t_{\mathrm{op}}\gg 0$ appearing in Eq.~(\ref{eq:method_ratio}) 
is usually not satisfied in practice, the excited states contaminations 
{could not be fully eliminated by tuning smearing parameters. For the purpose of eliminating the systematic uncertainties,
one should calculate the ratio (\ref{eq:method_ratio}) with several choices of $t_{\mathrm{sep}}$, and then makes sure whether the evaluated value of the nucleon matrix element does not change with a variation of $t_{\mathrm{sep}}$ within a certain precision. This is called the ratio method that is mainly used
in this study. 

In order to compare with the experimental values or other lattice results, the {\it bare matrix elements} obtained from the above mentioned method should be renormalized with the renormalization
constants $Z_{O_{\Gamma}}$ for each $\Gamma$ operator in a certain scheme. 
The Rome-Southampton method, which is known as the Regularization Independent (RI) scheme, 
is often used as the intermediate scheme in order to evaluate the renormalization constants $Z_{O_{\Gamma}}$ 
in fully nonperturbative manner. The resulting renormalization constants are then converted to the $\overline{\rm MS}$ 
scheme at certain scale $\mu_0$ and evolved to the scale of 2 GeV using the perturbation theory. 

In general, the final result of $Z^{\overline{\rm MS}}_{O_{\Gamma}}({\rm 2 GeV})$ receives 
the residual dependence of the choice of the matching scale $\mu_0$.
The perturbative conversion from the RI scheme to the $\overline{\rm MS}$ scheme produces 
the residual $\mu_0$ dependence. There are two main sources as follows.
One stems from lattice discretization errors at higher $\mu_0$, while another is originated from the nonperturbative 
effect that becomes relevant at lower $\mu_0$~\cite{Aoki2008}. 
The latter becomes serious when the physical lattice volume gets larger. However, as pointed out in Ref.~\cite{Aoki2008}, 
if a Symmetric MOMentum subtraction point is adopted in the RI scheme (hereafter regarded as the RI/SMOM scheme), the infrared effect 
is highly suppressed. In this paper, we use the RI/SMOM scheme for this purpose.
In order to reduce the systematic uncertainties associated with the residual $\mu_0$-dependence,
we used following two types of fitting functional forms as functions of the matching scale $\mu_0$
\begin{align}
    f_{O_{\Gamma}}^{\mathrm{Global}}(\mu_0) = \frac{c_{-1}}{(\Lambda^{-1}\mu_0)^2} + c_0 + 
    \sum_{k > 0}^{k_{\rm max}} c_k (a\mu_0)^{2k}
\quad \mathrm{and}\quad
    f_{O_{\Gamma}}^{\mathrm{IR}-\mathrm{trunc.}}(\mu_0) =  c_0 + \sum_{k > 0}^{k_{\rm max}}  c_k (a\mu_0)^{2k}
\label{eq:ren_fit}
\end{align}
with $c_0$ being the $\mu_0$-independent value of $Z^{\overline{\rm MS}}_{O_{\Gamma}}(\mu)$
at the renormalization scale $\mu=2$ GeV. 
The former functional form includes the term of the negative power of $\mu_0$,
which is induced by the nonperturbative effect characterized by a scale $\Lambda$. Therefore, the former is 
applied for fitting all data (denoted as ``Global'' in superscript), while 
the latter functional form is used for fitting the data in a restricted range of $\mu_0 \ge \mu$ 
(denoted as ``IR-trunc.'' in superscript).
The value of $k_{\rm max}$ is determined by a $\chi^2$ test for goodness fit, so that $k_{\rm max}=1$ and $2$ are chosen
in each case. The discrepancy on the values of $c_0$ extracted from 
these fittings with $k_{\rm max}=1, 2$ is quoted as a systematic error on the renormalization constant.

\section{Simulation details}
We mainly use the PACS10 configurations generated by the PACS Collaboration with the six stout-smeared ${\mathcal{O}}(a)$ improved Wilson-clover quark action and Iwasaki gauge action at $\beta=1.82$ corresponding to the lattice spacings of $0.085$ fm~\cite{Ishikawa2019-1,Ishikawa2019-2,Shintani2019,Taniguchi2012,Ishikawa:2021eut} with physical light quarks. 
When we compute nucleon two- and three-point functions, the all-mode-averaging (AMA) technique~\cite{Blum2013} is employed in order to reduce the statistical errors significantly without increasing
computational costs. Two lattice ensembles are generated with 
the same lattice spacing, but on 
different lattice sizes: $L^3\times T=128^3\times128$ and $64^3\times64$ corresponding $(10.9\ {\rm fm})^4$
and $(5.5\ {\rm fm})^4$ lattice volumes.
The smaller volume ensemble is used for the finite volume study on the  $g_A$ and nucleon elastic form factors, and also used for computing the renormalization constants which are known to be less sensitive to the finite volume effect.


\begin{table*}[ht]
    \caption{Details of the measurements: lattice size, time separation 
    ($t_{\rm sep}$), the smearing parameters $(A,B)$, the number of high-precision calculation ($N_{\mathrm{org}}$), the number of configurations ($N_{\mathrm{conf}}$), the measurements per configuration ($N_{G}$) and the total number of the  measurements
    ($N_{\mathrm{meas}}=N_{\mathrm{conf}} \times N_{G}$), respectively. 
\label{tab:measurements}}
\centering
\begin{tabular}{ccccccccc}
          \hline
Lattice size &  $t_{\rm sep}/a$ &Smearing parameters& $N_{\mathrm{org}}$&  $N_{G}$ & $N_{\mathrm{conf}}$ & $N_{\mathrm{meas}}$\\
\hline
          \hline
    $128^4$ lattice& 10&$(A,B)=(1.2,0.16)$ & 1& 128& 20& 2,560\\
                   & 12&             & 1& 256& 20& 5,120\\
                   & 14&             & 2& 320& 20& 6,400\\
                   & 16&             & 4& 512& 20& 10,240\\
          \hline
    $64^4$ lattice& 11& $(A,B)=(1.2,0.16)$ & 4& 40& 50& 2,000\\
                  & 14&                    & 4& 64& 100& 6,400\\
                   \cline{2-7}
                  & 12&$(A,B)=(1.2,0.14)$ & 4& 256& 100& 25,600\\
                  & 14&             & 4& 1,024& 100& 102,400\\
                  & 16&             & 4& 2,048& 100& 204,800\\
          \hline
\end{tabular}
\end{table*}

\section{Numerical results}
\black
In this study, we present the preliminary results for the renormalized values of the isovector axial, scalar
and tensor charges. All of the \textit{bare matrix elements} are evaluated with both $128^4$ and $64^4$ lattice ensembles. 
As for the renormalization, the local vector and axial currents are renormalized with the value of $Z_V=0.9513(76)(1487)$ 
and $Z_A=0.9650(68)(95)$ obtained by the Schr\"odinger functional scheme at the vanishing quark mass\cite{Ishikawa2016}, while 
the renormalization constants for the scalar and tensor charges are evaluated with RI/SMOM scheme as
described in Sec.\ref{sec:method}.

In Fig~\ref{fig:ren_axiscaten}, we show the $t_{\mathrm{sep}}$ dependence of $g_A$, $g_S$ and $g_T$, which are evaluated with the ratio method. The error bars for $g_A$ and the inner errors for $g_S$ and $g_T$ represent their statistical errors, while the outer ones represent the total error including the systematic error stemming from the renormalization. According to the analysis based on the ratio method, it is found that the condition $t_{\mathrm{sep}}>1$ (fm) is enough to suppress the excited states contaminations in all three channels. 

We compare our preliminary results of the renormalized values of $g_S$ (left) and $g_T$ (right) together with results from the recent lattice QCD calculations in Fig~\ref{fig:comparison}. 
Remark that our results are obtained solely from the physical point simulations which suffer 
from large statistical fluctuations, while the other lattice results are given by the combined data 
that includes the data taken from simulations at the heavier pion masses. However, the statistical and total errors 
on our results are comparable to the other lattice results, since we use the AMA method for the bare matrix elements
and the RI/SMOM scheme for the renormalization, both of which reduce the statistical and systematic uncertainties
on the final results. As for $g_S$, our preliminary result is consistent with the trend of the other results. 
On the other hand, our result of $g_T$ locates slightly higher than other continuum results, though 
this discrepancy would be caused by the discretization uncertainty that is not yet accounted in our calculations.

   \begin{figure}[tbh]
\centering
\includegraphics[width=0.325\linewidth,trim=0.0cm 1.0cm 2.5cm 2.5cm,clip]{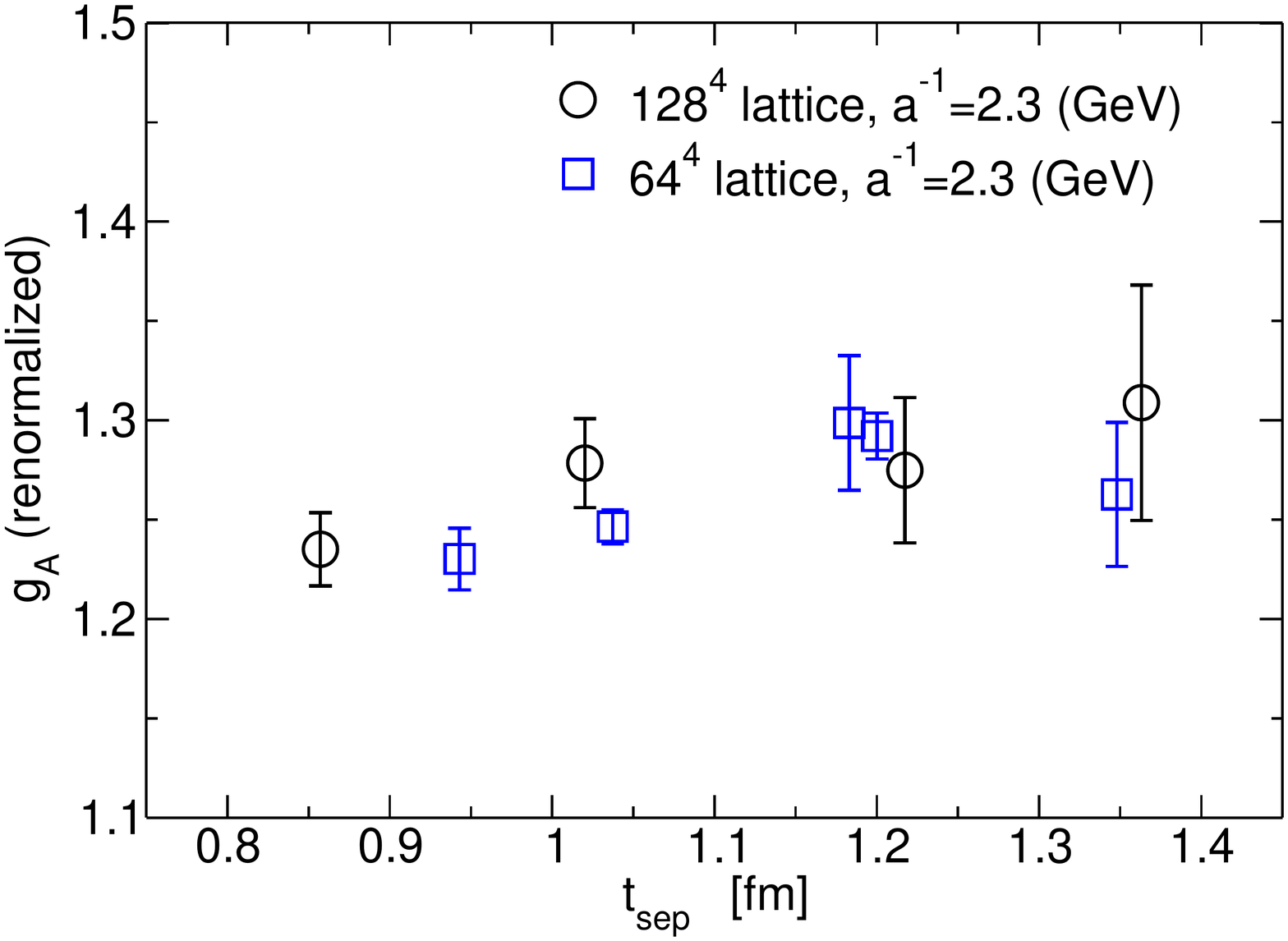}
\includegraphics[width=0.325\linewidth,trim=0.0cm 1.0cm 2.5cm 2.5cm,clip]{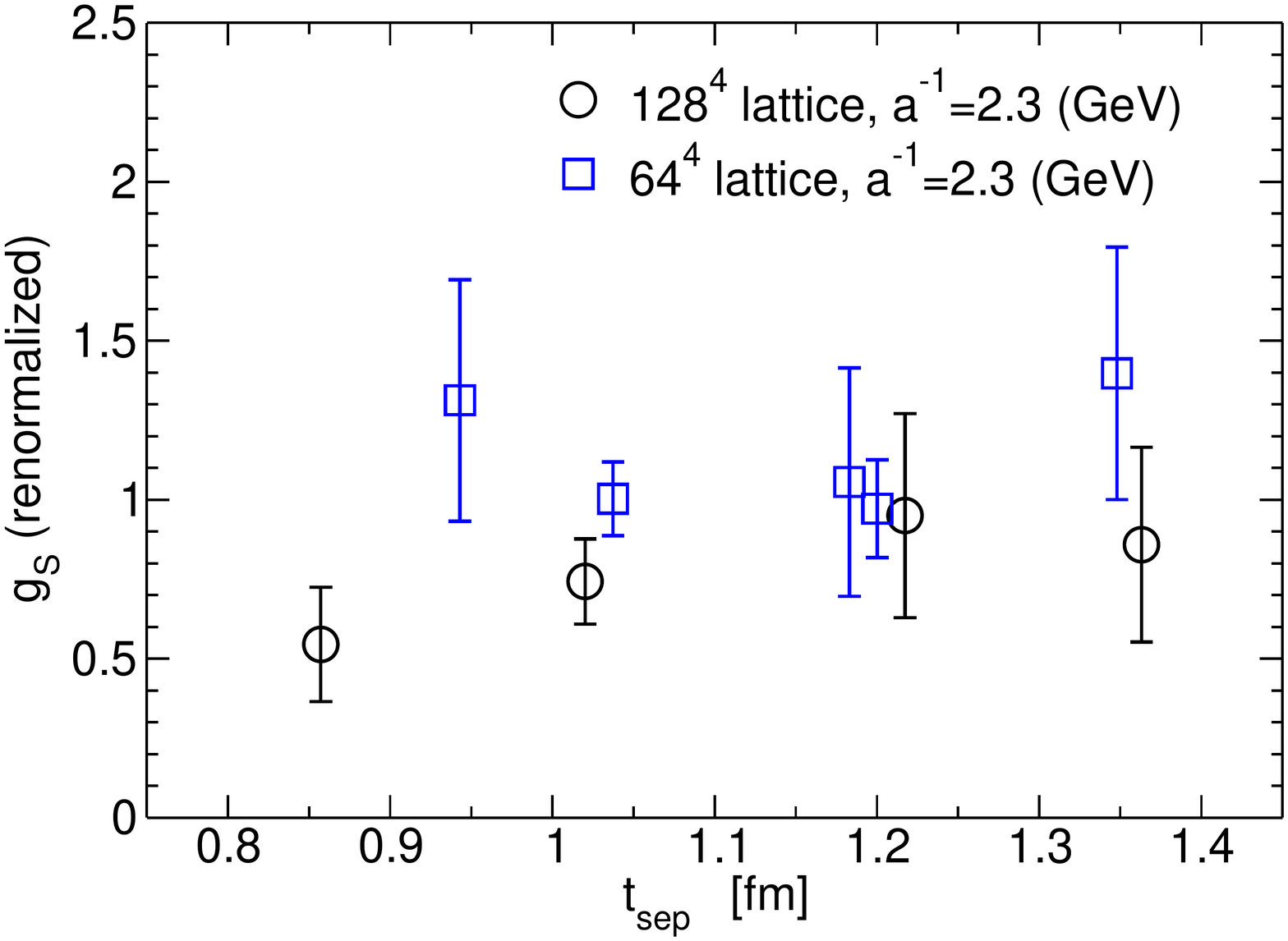}
\includegraphics[width=0.325\linewidth,trim=0.0cm 1.0cm 2.5cm 2.5cm,clip]{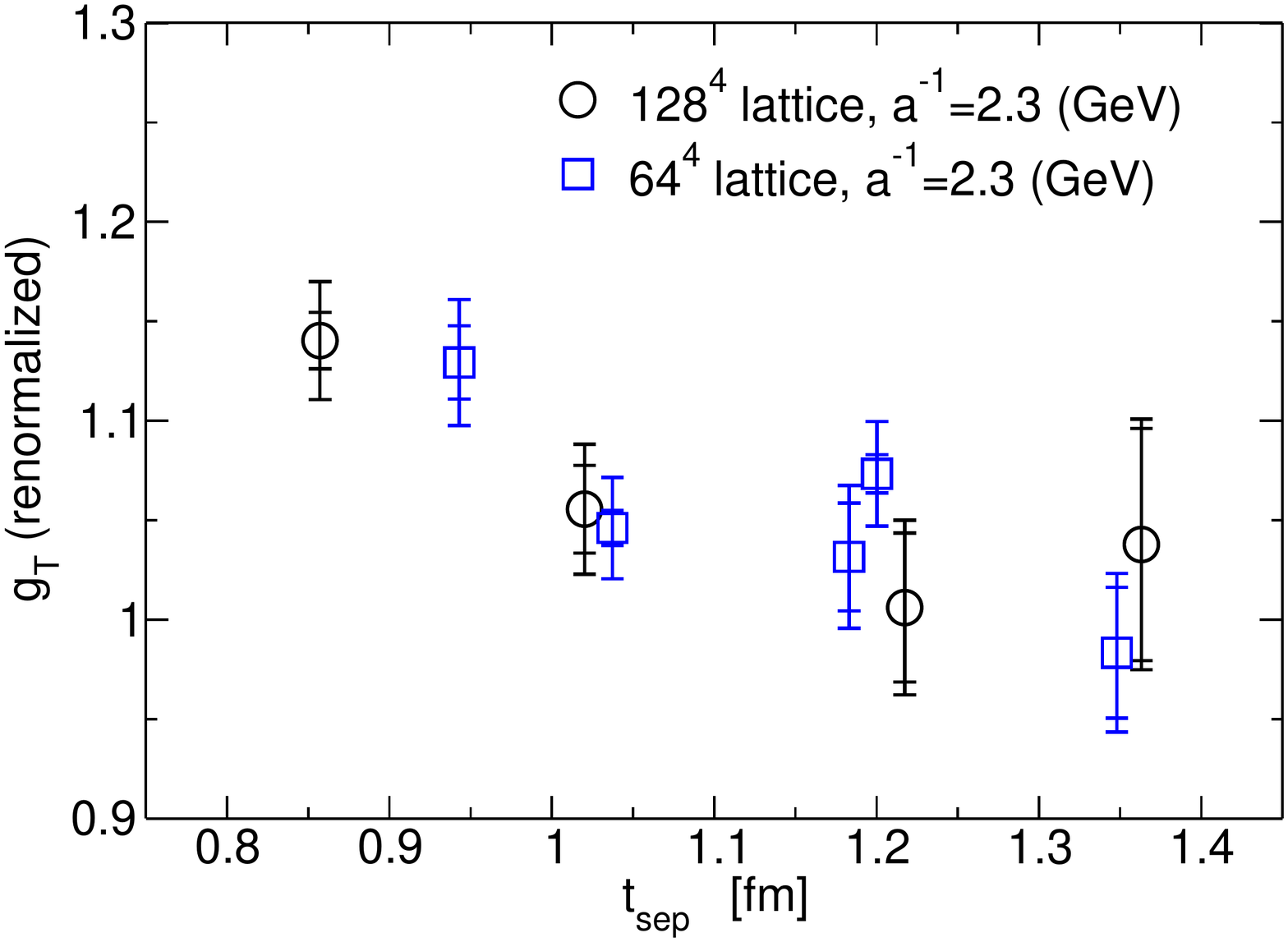}
\caption{
    $t_{\mathrm{sep}}$ dependence of the renormalized values of $g_A$ (left), $g_S$ (center) and $g_T$ (right). 
    The horizontal axis denotes the source-sink separation $t_{\mathrm{sep}}$ in physical unit. 
    The inner and outer error bars represent their statistical and total uncertainties, respectively.
\label{fig:ren_axiscaten}}
\end{figure}

\begin{figure}[tbh]
 \includegraphics[width=0.49\textwidth,trim= 2.3cm 0cm 9cm 0cm,clip]{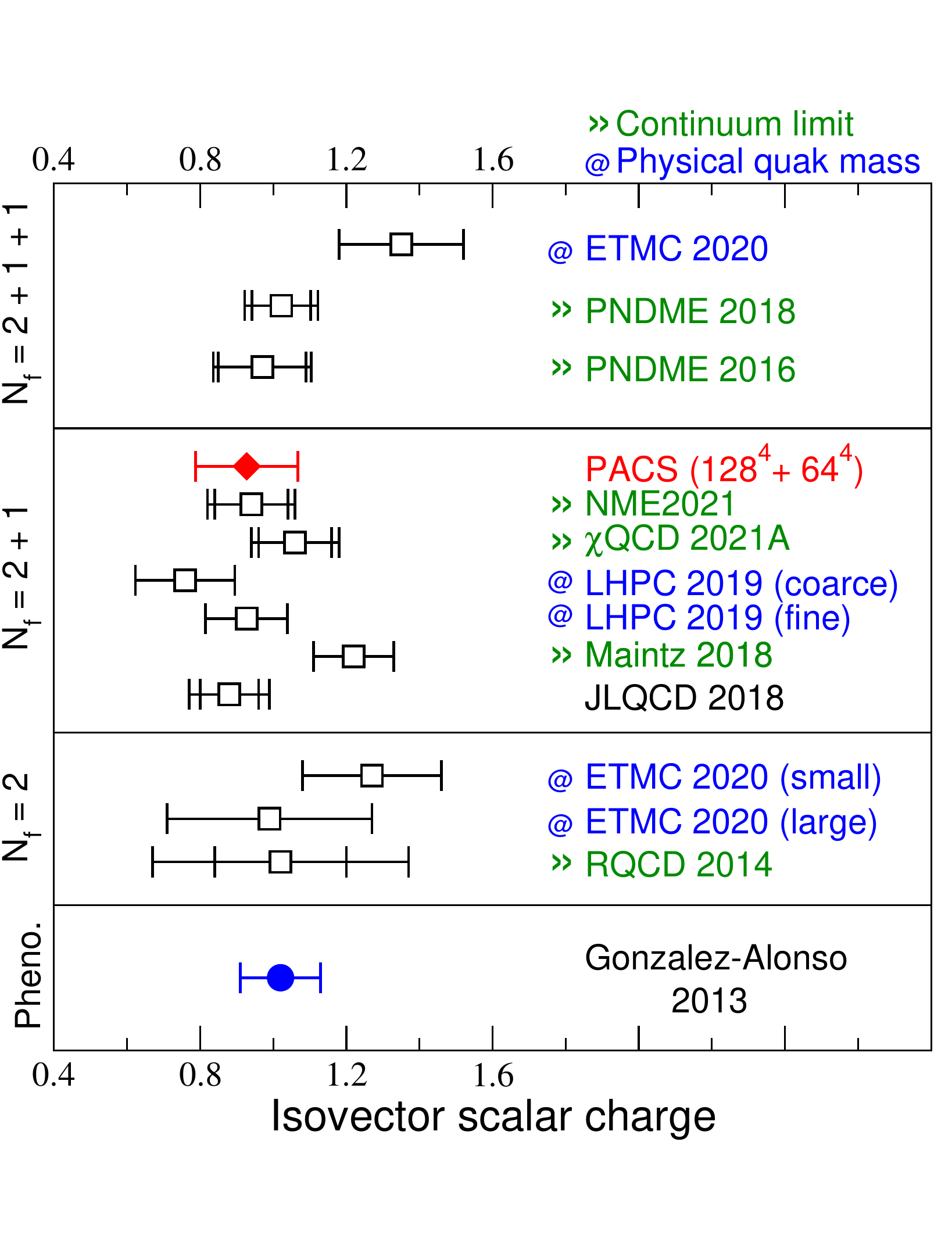}
 \includegraphics[width=0.49\textwidth,trim= 2.3cm 0cm 9cm 0cm,clip]{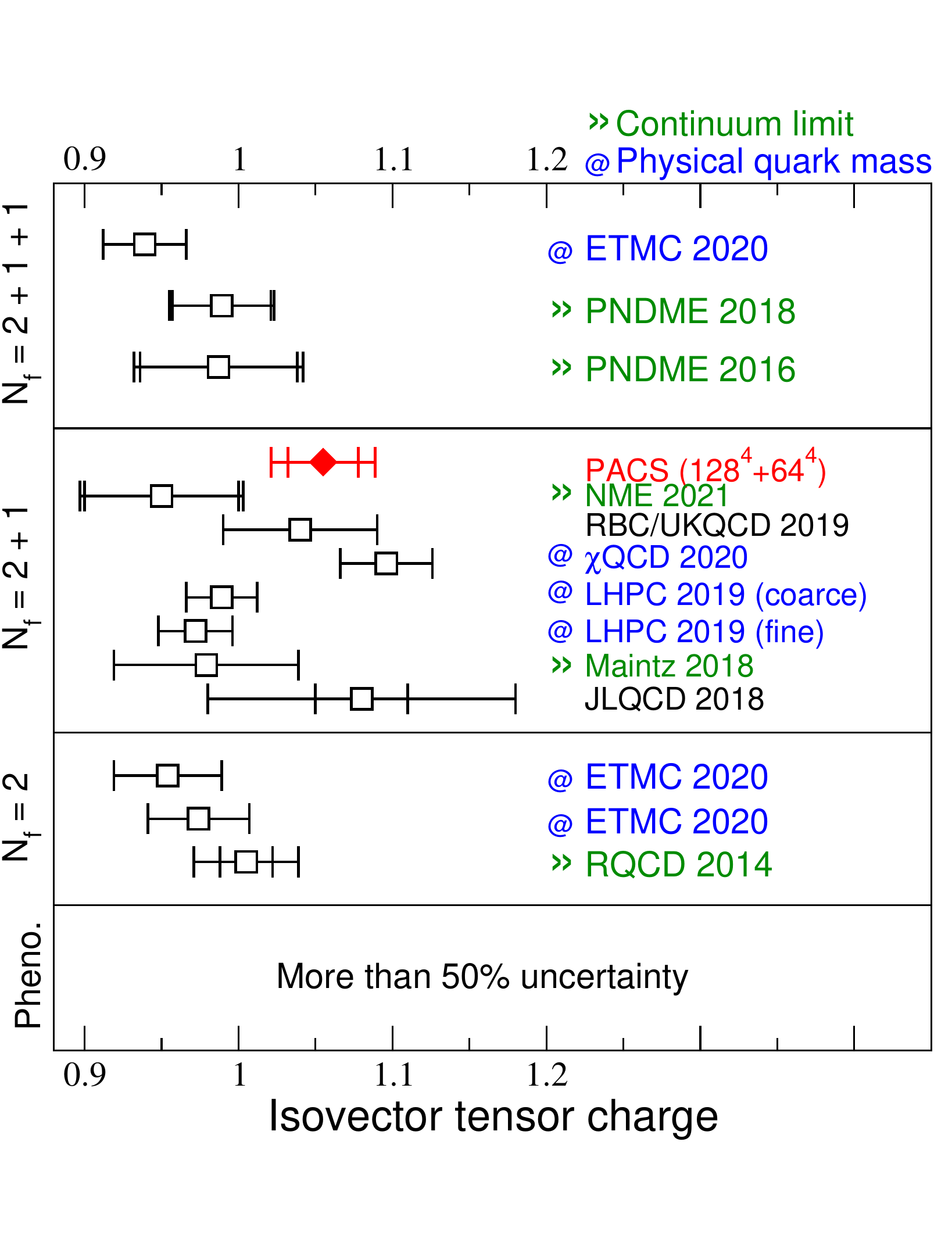}
 \caption{
 Comparison of our preliminary results (red diamonds) with the other lattice results (black squares) and the phenomenological value (blue circle)~\cite{Aoki:2021kgd,Gonzalez-Alonso:2013ura} for $g_S$ (left panel) and $g_T$ (right panel). 
The inner error bars represent the statistical uncertainties, while the outer ones represent the total uncertainties given 
by adding the statistical and systematic errors in quadrature. Blue labels indicate that the analysis includes 
the data from lattice QCD simulation near the physical point, while green labels indicate that the continuum extrapolation
is achieved.
}
 \label{fig:comparison}
\end{figure}

\section{Summary}
We have calculated the renormalized values of the nucleon isovector charge in the axial, scalar and tensor channels
using 2+1 flavor lattice QCD {\rm with physical light quarks}.  The calculations are carried out with the gauge configurations generated by the PACS Collaboration. In order to achieve high-precision and high-accuracy determination, we employ
the AMA technique which can reduce {the statistical error significantly, and the RI/SMOM scheme which keeps the systematic error under control. 
Consequently, we precisely determined $g_S$ and $g_T$ only at the physical point. 
Our results are comparable to the other lattice results, though the discretization uncertainties are not yet evaluated. 
Our research continues at a second, finer spacing ($a=0.064$ fm), and is now in progress\cite{Tsuji2021}.

\section*{Acknowledgement}
Numerical calculations in this work were performed on Oakforest-PACS in Joint Center for Advanced High Performance Computing (JCAHPC) and Cygnus in Center for Computational Sciences at University of Tsukuba under Multidisciplinary Cooperative Research Program of Center for Computational Sciences, University of Tsukuba, and Wisteria/BDEC-01 in the Information Technology Center, The University of Tokyo. This research also used computational resources through the HPCI System Research Projects (Project ID: hp170022, hp180051, hp180072, hp180126, hp190025, hp190081, hp200062, hp200188, hp210088) provided by Information Technology Center of the University of Tokyo and RIKEN Center for Computational Science (R-CCS). The  calculation employed OpenQCD system(http://luscher.web.cern.ch/luscher/openQCD/). This work was supported in part by Grants-in-Aid for Scientific Research from the Ministry of Education, Culture, Sports, Science and Technology(Nos. 18K03605, 19H01892). This work was supported by RIKEN Junior Research Associate Program.

\end{document}